\documentclass[a4paper,11pt]{article}

\usepackage{amsfonts,amssymb}
\usepackage{theorem}
\usepackage{epsfig}

\def\G{\Gamma}

\begin{document}

\rightline{IFUM-856-FT}
\vskip 0.8 truecm
\Large
\bf

\centerline{The Abelian Embedding Formulation of the St\"uckelberg Model}
\centerline{and its Power-counting Renormalizable Extension}

\normalsize
\rm

\vskip 0.5 truecm
\large
\centerline{Andrea Quadri \footnote{E-mail address: 
{\tt andrea.quadri@mi.infn.it}}}

\vskip 0.3 truecm
\normalsize
\centerline{Phys. Dept. University of Milan, 
via Celoria 16, 20133 Milan, Italy } 
\centerline{and I.N.F.N., sezione di Milano} 

\vskip 0.8 truecm
\centerline{\bf Abstract}

\begin{quotation}
\noindent
We elucidate the geometry of the polynomial formulation
of the non-abelian St\"uckel\-berg mechanism. 
We show that a natural off-shell nilpotent BRST
differential exists allowing to implement the
constraint on the $\sigma$ field by means of BRST techniques.
This is achieved by extending the ghost sector by an
 additional U(1) factor (abelian embedding).
An important consequence is that a further
BRST-invariant but not gauge-invariant mass term can be 
written for the non-abelian gauge fields.
As all versions of the St\"uckelberg theory, 
also the abelian embedding formulation yields a 
non power-counting renormalizable theory in $D=4$.
We then derive its natural power-counting renormalizable extension 
and show that the physical spectrum contains a
physical massive scalar particle. Physical unitarity is also
established. 
This  model implements the spontaneous symmetry breaking
in the abelian embedding formalism.
\end{quotation}

\newpage

\section{Introduction}\label{intro}

The St\"uckelberg formalism \cite{stueck,Ruegg:2003ps}
allows for a gauge-invariant mass term for non-abelian 
vector bosons without the need to introduce physical scalar
fields in the classical action. The main disadvantage 
of the non-abelian St\"uckelberg mechanism is the fact that it yields
a non power-counting renormalizable theory.  In particular 
the St\"uckelberg mass term 
\begin{eqnarray}
\frac{1}{2} m^2 {\rm Tr} [ (A_\mu - \frac{i}{g} \Omega \partial_\mu \Omega^\dagger ) ^2] \, , 
\label{intro:1}
\end{eqnarray}
with $\Omega = \exp (i g T_a \varphi_a(x))$ an element of the non-abelian
gauge group $G$, contains an infinite number of interaction vertices
involving the fields $\varphi_a(x)$.

There have been some attempts in the literature aiming
at a polynomial formulation of the St\"uckelberg mechanism
\cite{Dragon:1996tk,Slavnov:2005ip}. 
It is hoped that a polynomial interaction could help
in establishing a consistent subtraction scheme 
for the definition of the St\"uckelberg theory at the quantum level.

In Ref.~\cite{Slavnov:2005ip}
it has been pointed out that a polynomial action implementing
the St\"uckelberg construction can be derived from an interpolating
action which reproduces for different choices of its parameters
the St\"uckelberg theory, the Higgs model
as well as an embedding of the Higgs model which includes
additional physical scalar fields. The construction makes use
of a BRST-like (on-shell nilpotent) symmetry involving a pair
of ghost-antighost fields which are singlet under the non-abelian
gauge transformations.
An extension of this approach has been used in \cite{Slavnov:2006hz}
in order to propose  a model for massive gauge bosons without
fundamental scalars.

In this paper we elucidate the geometry of the polynomial
formulation of the St\"uckelberg theory. 
We show that the requirement of polynomiality
of the St\"uckel\-berg interaction can be formulated 
by means of  
a truly off-shell nilpotent BRST symmetry.
This leads to an abelian embedding
implementing the $\sigma$ model constraint 
by means of
an additional U(1) pair of ghost-antighost fields.
These fields play the
r\^ole of the $G$-singlet ghost-antighost fields 
proposed in \cite{Slavnov:2005ip}.
Moreover it turns out
that an abelian gauge connection $B_\mu$ can be introduced and given
a mass without violating the BRST invariance. $B_\mu$ can be 
chosen to be a free massive U(1) field.

The BRST invariants of this theory are particularly interesting
in the case of the group SU(2). For this group a polynomial
composite vector field can be constructed which transforms as a connection
under the BRST differential (but not under the SU(2) gauge transformations).
The rather surprising consequence is the possibility 
to generate a new polynomial BRST-invariant but not gauge-invariant
mass term for the non-abelian gauge fields. 

As all known versions of the St\"uckelberg mechanism, 
the abelian embedding model is not power-counting renormalizable.
We then study an extension thereof which is both power-counting renormalizable
and physically unitary.
Its physical spectrum is analyzed by BRST techniques and shown
to contain the three physical polarizations of the massive gauge fields
as well as a physical scalar particle. 
We prove by cohomological techniques 
that this theory is indeed physically unitary to all orders
in the perturbative expansion and give the whole set
of counterterms of the model.

This theory provides an alternative implementation of the
spontaneous symmetry breaking. 
Since it is power-counting renormalizable,
one could conjecture that it is physically equivalent
to the Higgs model, i.e. that it yields the same physical S-matrix elements.
The check of the  physical 
equivalence in the perturbative expansion is
an interesting question which deserves to be further investigated.

\medskip
The paper is organized as follows. In Sect.~\ref{sec.1} we briefly 
review
the standard formulation of the St\"uckelberg model based on
the use of a flat connection for the gauge group $G$ and discuss how the Higgs model
can be derived as the power-counting renormalizable extension
of the flat connection version of the St\"uckelberg theory.
In Sect.~\ref{sec.2} we develop the abelian embedding formalism for the
St\"uckelberg model. The additional BRST-invariant but not gauge-invariant
mass term that can be written for $G=\mbox{SU(2)}$ is discussed in Sect.~\ref{sec.3}.
In Sect.~\ref{sec.4} we move to the analysis of a  physically unitary 
and power-counting renormalizable extension of the abelian
embedding formalism. 
Power-counting renormalizability is established as a consequence of a set of functional
identities defining the theory.
The physical spectrum is constructed in Sect.~\ref{sec.6}.
Conclusions are finally given in Sect.~\ref{sec.7}.

\section{Flat connection formulation of the St\"uckelberg model}\label{sec.1}

For the sake of definiteness we consider the gauge group $G=\mbox{SU(2)}$. We follow
the derivation given in Ref.~\cite{Ferrari:2004yt} 
(for a review of the standard St\"uckelberg mechanism see also
\cite{Ruegg:2003ps}). The (global SU(2)-symmetric) Yang-Mills
action in the Proca gauge is
\begin{eqnarray}
S = \int d^4x \, \Big ( - \frac{1}{4} G_{a\mu\nu} G^{\mu\nu}_a + m^2 {\rm Tr} [A_\mu A^\mu] \Big ) 
\label{sec.1:1}
\end{eqnarray}
where $A_\mu = \tau_a A_{a\mu}$. $\tau_a$ are the Pauli matrices and $G_{a\mu\nu}$ is the
field strength
\begin{eqnarray}
G_{a \mu \nu} = \partial_\mu A_{a\nu} - \partial_\nu A_{a\mu} + g f_{abc} A_{b \mu} A_{c\nu} 
\label{sec.1:2}
\end{eqnarray}
with $f_{abc} = 2 \epsilon_{abc}$.
Let us now perform an operator-valued SU(2) local transformation
\begin{eqnarray}
A'_\mu = \Omega^\dagger A_\mu \Omega + \frac{i}{g} \Omega^\dagger \partial_\mu \Omega 
\label{sec.1:3}
\end{eqnarray}
with $\Omega \in \mbox{SU(2)}$. Then one gets the St\"uckelberg action
\begin{eqnarray}
&& 
\!\!\!\!\!\!\!\!\!\!\!\!\!\!\!\!\!\!\!\!
S = \int d^4x \, \Big ( - \frac{1}{4} G_{a\mu\nu} G^{\mu\nu}_a \nonumber \\
&& ~~~~~~ + \frac{m^2}{g^2} {\rm Tr} [ ( g \Omega^\dagger A_\mu \Omega + i \Omega^\dagger \partial_\mu \Omega )
           ( g \Omega^\dagger A_\mu \Omega + i \Omega^\dagger \partial_\mu \Omega ) ]  \Big ) \, .
\label{sec.1:4}
\end{eqnarray}
$S$ is invariant under the local SU(2) left transformations
\begin{eqnarray}
&& A'_\mu = U_L A_\mu U_L^\dagger + \frac{i}{g} U_L \partial_\mu U_L^\dagger \, , \nonumber \\
&& \Omega' = U_L \Omega \, .
\label{sec.1:5}
\end{eqnarray}
$\Omega$ is the St\"uckelberg field \cite{stueck,Ruegg:2003ps}.
The matrix $\Omega$ can be parameterized in terms of three independent fields $\phi_a$ as follows:
\begin{eqnarray}
\Omega = \frac{g}{2m} \Big ( \phi_0 \cdot 1 + i \tau_a \phi_a  \Big )
\label{sec.1:6}
\end{eqnarray}
with the constraint
\begin{eqnarray}
\phi_0^2 + \phi_a^2 = \frac{4m^2}{g^2} \, .
\label{sec.1:7}
\end{eqnarray}
Eq.(\ref{sec.1:7}) allows to express $\phi_0$ in terms of the fields $\phi_a$
\begin{eqnarray}
\phi_0 = \sqrt{\frac{4m^2}{g^2} - \phi_a^2} \, .
\label{sec.1:8}
\end{eqnarray}
Therefore, as a consequence of eq.(\ref{sec.1:8}),
the action $S$ in eq.(\ref{sec.1:4}) contains an infinite number 
of interaction vertices and the theory is
not renormalizable by power-counting (in $D=4$).
Physical unitarity of the St\"uckelberg model in the Landau gauge has been discussed in detail 
in \cite{Ferrari:2004pd}.

By setting
\begin{eqnarray}
&& \Phi = \frac{\sqrt{2} m}{g} ~ \Omega v_+ = \frac{1}{\sqrt{2}} \pmatrix{i \phi_1 + \phi_2 \cr \phi_0 - i \phi_3} \,  , 
\label{sec.1:10}
\end{eqnarray}
with $v_+^T = \pmatrix{ 0 & 1}$
the St\"uckelberg mass term reduces to
\begin{eqnarray}
\int d^4 x \, (D_\mu \Phi)^\dagger (D^\mu \Phi) \, .
\label{sec.1:12}
\end{eqnarray}
By dropping the constraint on the field $\phi_0$ one obtains
the Higgs model \cite{Higgs:1964ia,Kibble:1967sv,'tHooft:1971rn}. 
In contrast with the St\"uckelberg model, the Higgs model is power-counting renormalizable
in $D=4$.
In the Higgs model
$\phi_0$ becomes an independent field. 
As is well-known, in addition to the gauge-invariant term in eq.(\ref{sec.1:12})
power-counting renormalizability in $D=4$ allows for two further invariants depending on $\Phi$, namely
$\int d^4x \, \Phi^\dagger \Phi$ and $\int d^4x \, (\Phi^\dagger \Phi)^2$.
Their coefficients can be chosen in such a way that spontaneous symmetry breaking is triggered 
by the tree-level potential and consequently
$\phi_0$ acquires a non-vanishing v.e.v. $v$. The resulting action depends on an additional parameter $\lambda$
which controls the strength of the quartic Higgs self-interaction:
\begin{eqnarray}
S_{\lambda} = \int d^4x \, \Big ( - \frac{1}{4} G_{a\mu\nu} G^{\mu\nu}_a
+ (D_\mu \Phi)^\dagger (D^\mu \Phi) - \lambda (\Phi^\dagger \Phi - \frac{v^2}{2} )^2 \Big ) \, .
\label{sec.1:14.bis}
\end{eqnarray}
After the shift $\phi_0 = \sigma + v$, which amounts to the redefinition
\begin{eqnarray}
\Phi \rightarrow \Phi + \frac{v}{\sqrt{2}} v_+ \, ,
\label{sec.1:15}
\end{eqnarray}
one obtains a theory for massive non-abelian gauge bosons which contains an additional physical scalar
particle described by the physical Higgs field $\sigma$. 
The St\"uckelberg model can be formally obtained
by taking the limit $\lambda \rightarrow \infty$ in the action (\ref{sec.1:14.bis}), yielding the constraint
\begin{eqnarray}
\Phi^\dagger \Phi - \frac{v^2}{2} = 0 \, .
\label{sec.1:15.bis}
\end{eqnarray}
This coincides with eq.(\ref{sec.1:7}) by setting $m = \frac{gv}{2}$.

\section{Abelian embedding formulation of the St\"uckelberg model}\label{sec.2}

In Ref.~\cite{Slavnov:2005ip} it has been pointed out that a polynomial 
action implementing the St\"uckelberg 
mechanism can be derived from an interpolating
action which reproduces for different choices of its parameters the St\"uckelberg theory,
the Higgs  model as well as an embedding of the Higgs 
model which includes
additional physical scalar fields.  
The construction makes use of a BRST-like (on-shell nilpotent) symmetry involving
a pair of ghost-antighost fields which are singlet under the non-abelian
gauge transformations.

It is the purpose of this Section to obtain a polynomial formulation of the St\"uckelberg action
based on a truly off-shell nilpotent BRST symmetry.
We perform a $R_\xi$-gauge-fixing of the action 
\begin{eqnarray}
S_0 = \int d^4x \, \Big ( - \frac{1}{4} G_{a\mu\nu} G_a^{\mu\nu} 
+ (D_\mu \Phi)^\dagger D^\mu \Phi \Big ) 
\label{sec.2:0}
\end{eqnarray}
in the BRST formalism and obtain the gauge-fixed action
\begin{eqnarray}
S'_0 & = & S_0 + \int d^4x \Big ( \frac{\xi}{2} B_a^2 - B_a ( \partial A_a + \xi gv \phi_a) \nonumber \\
    &   & + \bar \omega_a [ \partial^\mu (D_\mu \omega)_a + \xi g^2 v (\sigma + v) \omega_a + \xi g^2 v 
                \epsilon_{abc} \phi_b \omega_c ] \Big ) \, .
\label{sec.2:1}
\end{eqnarray}                
$\omega_a$ are the non-abelian ghost fields, $\bar \omega_a$ the corresponding antighosts. $B_a$ are the
Nakanishi-Lautrup multiplier fields. $\xi$ is the gauge parameter. 

$S'_0$ is invariant 
under the following BRST differential
\begin{eqnarray}
&& s A_{a \mu} = (D_\mu \omega)_a = \partial_\mu \omega_a + g f_{abc} A_{b\mu} \omega_c \, , \nonumber \\
&& s \omega_a = - \frac{1}{2} g f_{abc} \omega_b \omega_c \, ,  \nonumber \\
&& s \Phi = i g \omega_a \tau_a \Phi \,  , ~~~~ s \phi_0 = -g \omega_a \phi_a \, , ~~~~
s \phi_a = g (\omega_a \phi_0 + \epsilon_{abc} \phi_b \omega_c) \, , \nonumber \\
&& s \bar \omega_a = B_a \, , ~~~~ s B_a = 0 \, .
\label{sec.2:2}
\end{eqnarray}
At this point we wish to implement the constraint in eq.(\ref{sec.1:15.bis}) 
\begin{eqnarray}
\Phi^\dagger \Phi - \frac{v^2}{2} = \frac{1}{2} \sigma^2 +  v \sigma + \frac{1}{2} \phi_a^2 = 0 
\label{sec.2:3}
\end{eqnarray}
by means of BRST techniques. The simplest possibility
is to introduce an antighost field $\bar c$ transforming under $s$ as follows:
\begin{eqnarray}
s \bar c = \Phi^\dagger \Phi - \frac{v^2}{2} \, .
\label{sec.2:4}
\end{eqnarray}
Since the constraint in eq.(\ref{sec.2:3}) is gauge-invariant, $s^2 \bar c=0$.
One should also introduce the ghost $c$ corresponding to the antighost $\bar c$, which we pair 
in a BRST doublet \cite{Piguet:1995er,Barnich:2000zw,Quadri:2002nh} with
a scalar field $X$ as follows
\begin{eqnarray}
s X = v c \, , ~~~~ s c =0 \, .
\label{sec.2:5}
\end{eqnarray}
Although it is not strictly necessary, it is tempting to consider 
$c$ as the abelian ghost of a U(1) connection $B_\mu$, so that one might also 
set
\begin{eqnarray}
s B_\mu = \partial_\mu c \, .
\label{sec.2:6}
\end{eqnarray}
Then the original BRST symmetry is embedded in a larger differential with an abelian component 
given by eqs.(\ref{sec.2:5}) and (\ref{sec.2:6}).

We remark that in the embedding theory the quartic potential in $S_\lambda$ 
in eq.(\ref{sec.1:14.bis}) is $s$-exact since
\begin{eqnarray}
- \int d^4x \,  \lambda (\Phi^\dagger \Phi - \frac{v^2}{2} )^2 = s \Big [ \int d^4x \, \Big ( -\lambda \bar c 
~ (\Phi^\dagger \Phi - \frac{v^2}{2} ) \Big ) \Big ] \, .
\label{sec.2:6.bis}
\end{eqnarray}

By adding to the action $S'_0$ in eq.(\ref{sec.2:1}) the following term
\begin{eqnarray}
S_{constr} & = & \int d^4x \, s (\frac{1}{v} X \square \bar c) \nonumber \\
           & = & \int d^4x \, \Big [ - \bar c \square c + \frac{1}{2v} X \square (  \sigma^2 + 2 v \sigma + \phi_a^2 ) \Big ] 
\label{sec.2:7}
\end{eqnarray}
the constraint of the St\"uckelberg model is reproduced in the way suggested in
Ref.~\cite{Slavnov:2005ip}. The non-renormalizability by power-counting 
is induced by the interaction vertices $\frac{1}{2v} X \square (\sigma^2 + \phi_a^2)$ in the R.H.S. of eq.(\ref{sec.2:7}).

By looking at eqs.(\ref{sec.2:5}) and (\ref{sec.2:6}) it is also clear that one can add a 
kinetic term and a mass term for $B_\mu$ without violating the BRST symmetry:
\begin{eqnarray}
S_{\mbox{U(1)}} = \int d^4x \, \Big ( - \frac{1}{4} F_{\mu\nu} F^{\mu\nu}
+ \frac{1}{2} M^2 \Big ( B_\mu - \frac{1}{v} \partial_\mu X \Big )^2 \Big ) \, ,
\label{sec.2:8}
\end{eqnarray}
where
\begin{eqnarray}
F_{\mu\nu} = \partial_\mu B_\nu - \partial_\nu B_\mu \, .
\label{sec.2:10}
\end{eqnarray}
With the choice of eq.(\ref{sec.2:8}) $B_\mu$ is a free massive U(1) gauge field. 

\section{Further BRST-invariant mass terms}\label{sec.3}

We now go back to eq.(\ref{sec.2:0}) 
and consider the current linearly coupled to 
the gauge fields $A_{a\mu}$:
\begin{eqnarray}
j^\mu_a = -ig \partial^\mu \Phi^\dagger \tau_a \Phi + ig \Phi^\dagger \tau_a 
\partial^\mu \Phi \, .
\label{sec2.1}
\end{eqnarray}
We evaluate its variation under a gauge transformation $\delta \Phi = i g \alpha_a \tau_a \Phi$ with gauge parameters $\alpha_a(x)$:
\begin{eqnarray}
\delta j^\mu_a & = & -2 g^2 \Phi^\dagger \Phi \partial^\mu \alpha_a + 
g f_{abc} j^\mu_b \alpha_c \, . 
\label{sec2.2}
\end{eqnarray}
This is not the transformation of a gauge connection due to the appearance
of the factor $\Phi^\dagger \Phi$ in front of the gradient of $\alpha_a$.
This factor can be compensated by the abelian antighost field as follows.
We consider the composite vector field 
\begin{eqnarray}
{\tilde F}^\mu_a = j^\mu_a + 2g^2 \bar c \partial^\mu \omega_a \, 
\label{sec2.3}
\end{eqnarray}
and compute its BRST variation:
\begin{eqnarray}
s {\tilde F}^\mu_a & = & s j^\mu_a + 2g^2 \Big ( \Phi^\dagger \Phi - \frac{v^2}{2} \Big ) \partial^\mu \omega_a -2g^2 \bar c \partial^\mu (-\frac{g}{2}
f_{abc} \omega_b \omega_c ) \nonumber \\
          & = & -2 g^2 \Phi^\dagger\Phi \partial^\mu \omega_a 
                + g f_{abc} j^\mu_b \omega_c +
                2 g^2 \Big ( \Phi^\dagger \Phi - \frac{v^2}{2} \Big )                           \partial^\mu \omega_a \nonumber \\
          &   & + 2 g^2 \bar c ~ g f_{abc} \partial^\mu \omega_b \omega_c
                \nonumber \\
          & = & -v^2 g^2 \partial^\mu \omega_a + g f_{abc} ({\tilde F}^\mu_b 
                -2g^2 \bar c  \partial^\mu \omega_b) \omega_c 
                \nonumber \\
          &   & + 2 g^2 \bar c ~ g f_{abc} \partial^\mu \omega_b \omega_c
                \nonumber \\
          & = & - v^2 g^2 \partial^\mu \omega_a + g f_{abc}{\tilde F}^\mu_b \omega_c \, .    
\label{sec2.4}
\end{eqnarray}
The above equation allows us to derive a vector field
which transforms as a connection under $s$ 
by properly rescaling $\tilde F^\mu_a$:
by setting 
\begin{eqnarray}
F^\mu_a & = & - \frac{1}{g^2 v^2} \tilde F^\mu_a \nonumber \\
        & = & - \frac{1}{g^2 v^2} 
                \Big (
                   -i g \partial^\mu \Phi^\dagger \tau_a \Phi
                   +i g \Phi^\dagger \tau_a \partial^\mu \Phi
                   +2 g^2 \bar c \partial^\mu \omega_a 
                \Big )    
\label{sec2.5}
\end{eqnarray}
we get
\begin{eqnarray}
s F^\mu_a = \partial^\mu \omega_a + g f_{abc} F^\mu_b \omega_c \, .
\label{sec2.6}
\end{eqnarray}
By eq.(\ref{sec2.6}) one can  use $F^\mu_a$ in order to generate a new polynomial BRST-invariant (but not gauge-invariant) mass term for $A^\mu_a$, given by
\begin{eqnarray}
\frac{1}{2} m^2 (A_a^\mu - F^\mu_a)^2 \, .
\label{sec2.7}
\end{eqnarray}
This term is absent in the standard 
flat connection formulation of the St\"uckelberg theory.

As a final point we remark that for an arbitrary gauge group $G$ 
with generators $T_a$ eq.(\ref{sec2.2}) becomes
\begin{eqnarray}
\delta j^\mu_a & = & - g^2 \Phi^\dagger \{ T_a, T_b \} \Phi \partial^\mu \alpha_b + 
g f_{abc} j^\mu_b \alpha_c \, . 
\label{sec2.8}
\end{eqnarray}
From the above equation we see that in order to apply the compensation mechanism
based on the abelian antighost $\bar c$ the anticommutator $\{ T_a, T_b \}$ 
has to be proportional to $\delta_{ab}$ times the identity matrix. 

\section{A power-counting renormalizable extension of the abelian embedded St\"uckelberg model}
\label{sec.4}

In this Section we discuss a mechanism for obtaining a power-counting renormalizable
theory of massive gauge bosons from the abelian embedding formulation of the
St\"uckelberg model.
We will require that the BRST differential
controlling the theory is off-shell nilpotent. 
Moreover we wish to formulate the theory without higher derivatives.
For that purpose we set now
\begin{eqnarray}
&& sX_1 = vc \, , ~~~~ sc =0 \, , \nonumber \\
&& s \bar c = \Phi^\dagger \Phi - \frac{v^2}{2} -v X_2 \, , ~~~~ s X_2 = 0 \, .
\label{sec.3:1}
\end{eqnarray}
The gauge-fixed action obtained from $S_\lambda$ is
\begin{eqnarray}
S'_\lambda & = & S_\lambda + \int d^4x \Big ( \frac{\xi}{2} B_a^2 - B_a ( \partial A_a + \xi gv \phi_a) 
          \nonumber \\
    &   & + \bar \omega_a [ \partial^\mu (D_\mu \omega)_a + \xi g^2 v (\sigma + v) \omega_a + \xi g^2 v 
                \epsilon_{abc} \phi_b \omega_c ] \Big ) \, .
\label{sec.3:2}
\end{eqnarray}
To this action we add 
\begin{eqnarray}
\!\!\!\!\!\!\!\!\!\!\!
S_{constr,X_2} & = & \int d^4x \, \Big [ s \Big ( \frac{X_1 + X_2}{v} (\square + 2 \lambda v^2) \bar c \Big ) - \frac{v^2}{2\rho} X_2^2 \Big ] \nonumber \\
             & = & \int d^4x \, \Big ( - \bar c (\square + 2 \lambda v^2) c \nonumber \\
             & & +  \frac{1}{v} (X_1 + X_2) (\square + 2 \lambda v^2) ( \frac{1}{2} \sigma^2 + v \sigma + \frac{1}{2} \phi_a^2 - v X_2 ) - \frac{v^2}{2\rho} X_2^2 \Big ) \, . \nonumber \\
\label{sec.3:3}
\end{eqnarray}
We also introduce the antifields for $A_{a\mu}, \omega_a, \sigma, \phi_a, \bar c$, which we denote
by $A^*_{a\mu}, \omega_a^*, \sigma^*, \phi_a^*, \bar c^*$. The complete action is finally given by
\begin{eqnarray}
\G^{(0)} & = & S_\lambda' + S_{constr,X_2} \nonumber \\
         &   & + \int d^4x \, \Big [ A_{a\mu}^* (D^\mu \omega)_a - g \sigma^* \omega_a \phi_a 
                                    + g \phi_a^* (\omega_a (\sigma + v) + \epsilon_{abc} \phi_b \omega_c) 
                                     \nonumber \\
         &   &  ~~~~~~~~~ -\frac{1}{2} \omega_a^* ~ g f_{abc} \omega_b \omega_c + \bar c^* ( \frac{1}{2} \sigma^2 + v \sigma + \frac{1}{2} \phi_a^2 - vX_2) \Big ] \, . \nonumber \\
\label{sec.3:4}
\end{eqnarray}
We can assign a ghost number to the fields and antifields of the theory.
$A_{a\mu},\phi_a,\sigma,B_a,X_1,X_2, \bar c^*$ have ghost number zero,
$A_{a\mu}^*,\phi_a^*, \sigma^*, \bar \omega_a, \bar c$ have ghost number $-1$,
$\omega_a$ and $c$ ghost number $+1$ while $\omega_a^*$ has ghost number $-2$.

We notice that 
$\G^{(0)}$ is separately invariant under the BRST differential $s_0$ given by
\begin{eqnarray}
&& s_0 A_{a \mu} = (D_\mu \omega)_a = \partial_\mu \omega_a + g f_{abc} A_{b\mu} \omega_c \, , \nonumber \\
&& s_0 \omega_a = - \frac{1}{2} g f_{abc} \omega_b \omega_c \, ,  \nonumber \\
&& s_0 \sigma = -g \omega_a \phi_a \, , ~~~~
s_0\phi_a = g (\omega_a (\sigma + v) + \epsilon_{abc} \phi_b \omega_c) \, , \nonumber \\
&& s_0 \bar \omega_a = B_a \, , ~~~~ s_0 B_a = 0 \, , ~~~~ s_0 \bar c = s_0 c = s_0 X_1 = s_0 X_2 = 0 
\label{sec.3:BRST1}
\end{eqnarray}
and under the BRST differential $s_1$ given by
\begin{eqnarray}
&& s_1 X_1 = vc \, , ~~~~ s_1 c =0 \, , \nonumber \\
&& s_1 \bar c = \Phi^\dagger \Phi - \frac{v^2}{2} -v X_2 \, , ~~~~ s_1 X_2 = 0 \, , \nonumber \\
&& s_1 A_{a\mu} = s_1 \sigma = s_1 \phi_a =  s_1 \omega_a = s_1 \bar \omega_a= s_1 B_a = 0 \, .
\label{sec.3:BRST2}
\end{eqnarray}
$\G^{(0)}$ fulfills the following functional identities
\begin{itemize}
\item the non-abelian ghost equation
\begin{eqnarray}
&& \frac{\delta \G^{(0)}}{\delta \bar \omega_a} = \partial^\mu \frac{\delta \G^{(0)}}{\delta A^{\mu*}_a}
   + \xi gv \frac{\delta \G^{(0)}}{\delta \phi_a^*} \,
\label{sec.3:5}
\end{eqnarray}
\item the abelian ghost equation
\begin{eqnarray}
&& \frac{\delta \G^{(0)}}{\delta \bar c} = - (\square + 2 \lambda v^2) c 
\label{sec.3:6}
\end{eqnarray}
\item the abelian antighost equation
\begin{eqnarray}
&& \frac{\delta \G^{(0)}}{\delta c} = (\square + 2 \lambda v^2) \bar c 
\label{sec.3:7}
\end{eqnarray}
\item the $B$-equation
\begin{eqnarray}
\frac{\delta \G^{(0)}}{\delta B_a} = \xi B_a - \partial A_a - \xi gv \phi_a 
\label{sec.3:8}
\end{eqnarray}
\item the $X_1$-equation
\begin{eqnarray}
\frac{\delta \G^{(0)}}{\delta X_1} = \frac{1}{v} (\square + 2 \lambda v^2) \frac{\delta \G^{(0)}}{\delta \bar c^*} 
\label{sec.3:9}
\end{eqnarray}
\item the $X_2$-equation
\begin{eqnarray}
\!\!\!\!\!
\frac{\delta \G^{(0)}}{\delta X_2} = \frac{1}{v} (\square + 2\lambda v^2) \frac{\delta \G^{(0)}}{\delta \bar c^*}  - (\square + 2 \lambda v^2) (X_1 + X_2) - \frac{v^2}{\rho} X_2 - v \bar c^*
\label{sec.3:10}
\end{eqnarray}
\item the Slavnov-Taylor (ST) identity
\begin{eqnarray}
\!\!\!\!\!\! {\cal S}(\G^{(0)}) & = & \int d^4x \, 
                   \Big ( \frac{\delta \G^{(0)}}{\delta A^{\mu*}_a} \frac{\delta \G^{(0)}}{\delta A_{a\mu}} 
                        + \frac{\delta \G^{(0)}}{\delta \sigma^*} \frac{\delta \G^{(0)}}{\delta \sigma}
                        + \frac{\delta \G^{(0)}}{\delta \phi^*_a} \frac{\delta \G^{(0)}}{\delta \phi_a} \nonumber \\                        &   & ~~~~~~ 
                        + \frac{\delta \G^{(0)}}{\delta \omega_a^*} \frac{\delta \G^{(0)}}{\delta \omega_a}
                        + \frac{\delta \G^{(0)}}{\delta \bar c^*} \frac{\delta \G^{(0)}}{\delta \bar c} 
                        + B_a \frac{\delta \G^{(0)}}{\delta \bar \omega_a} 
                        + v c \frac{\delta \G^{(0)}}{\delta X_1}
                   \Big )  = 0 \, .  \nonumber \\
\label{sec.3:11}
\end{eqnarray}
By virtue of eq.(\ref{sec.3:7}) invariance of $\G^{(0)}$ under $s_0$  is recovered 
by projecting  eq.(\ref{sec.3:11}) at order zero in powers of $c$ while invariance of $\G^{(0)}$ under $s_1$ 
is obtained  by projecting eq.(\ref{sec.3:11}) at order one in powers of $c$.

The choice of gathering both invariances into a single ST identity equipped with the grading in $c$
proves useful in the renormalization of the model, as is discussed in the Sect.~\ref{sec:cts}.

\end{itemize}

\subsection{Power-counting rules}

In Appendix~\ref{appA} we give the propagators of the model.
Diagonalization of the quadratic part in the fields of $\G^{(0)}$ is achieved by setting
\begin{eqnarray}
&& B'_a = B_a - \frac{1}{\xi} (\partial A_a + \xi gv \phi_a) \, , \nonumber \\
&& \sigma' = \sigma - X_1 - X_2 \, .
\label{sec.3:12}
\end{eqnarray}
The corresponding UV mass dimensions of the fields
and external sources can be summarized as follows. $A_{a\mu}, \sigma', X_1, X_2$, $\phi_a, \bar \omega_a$, $\omega_a$,
$\bar c$ and $c$ have dimension one, $B'_a$ has dimension two.
$\bar c^*$, $A_{a\mu}^*, \phi_a^*, \sigma^*$ and $\omega^*_a$ have dimension two.

All interaction vertices in $\G^{(0)}$ with the exception of 
\begin{eqnarray}
\frac{1}{v} (X_1 + X_2) \square ( \frac{1}{2} \sigma^2 + \frac{1}{2} \phi_a^2)
\label{sec.3:13}
\end{eqnarray}
have UV dimension $\leq 4$. We remark for future use 
that the interaction vertices
depend on $X_1, X_2$ only via the combination $X_1+X_2$.

\subsection{Power-counting renormalizability}\label{sec:ren}

In this Section we show that the model is indeed power-counting
renormalizable, despite the fact that it contains the vertices
in eq.(\ref{sec.3:13}). We impose eqs.(\ref{sec.3:5})-(\ref{sec.3:10})
on the 1-PI vertex functional $\G$:
\begin{eqnarray}
&& \frac{\delta \G}{\delta \bar \omega_a} = \partial^\mu \frac{\delta \G}{\delta A^{\mu*}_a}
   + \xi gv \frac{\delta \G}{\delta \phi_a^*} \, ,
\label{sec.3:e1}
\end{eqnarray}
%
%
\begin{eqnarray}
&& \frac{\delta \G}{\delta \bar c} = - (\square + 2 \lambda v^2) c \, , 
\label{sec.3:e2}
\end{eqnarray}
%
%
\begin{eqnarray}
&& \frac{\delta \G}{\delta c} = (\square + 2 \lambda v^2) \bar c  \, , 
\label{sec.3:e3}
\end{eqnarray}
%
%
\begin{eqnarray}
&& \frac{\delta \G}{\delta B_a} = \xi B_a - \partial A_a - \xi gv \phi_a \, ,  
\label{sec.3:e4}
\end{eqnarray}
%
%
\begin{eqnarray}
&& \frac{\delta \G}{\delta X_1} = \frac{1}{v} (\square + 2 \lambda v^2) \frac{\delta \G}{\delta \bar c^*} \, , 
\label{sec.3:e5}
\end{eqnarray}
%
%
\begin{eqnarray}
&& 
\!\!\!\!\!\!\!
\frac{\delta \G}{\delta X_2} = \frac{1}{v} (\square + 2\lambda v^2) \frac{\delta \G}{\delta \bar c^*}  -  (\square + 2 \lambda v^2) (X_1 + X_2) - \frac{v^2}{\rho} X_2 - v \bar c^* \, .
\label{sec.3:e6}
\end{eqnarray}
The above set of functional equations hold together with the ST identity
\begin{eqnarray}
\!\!\!\!\!\! {\cal S}(\G) & = & \int d^4x \, 
                   \Big ( \frac{\delta \G}{\delta A^{\mu*}_a} \frac{\delta \G}{\delta A_{a\mu}} 
                        + \frac{\delta \G}{\delta \sigma^*} \frac{\delta \G}{\delta \sigma}
                        + \frac{\delta \G}{\delta \phi^*_a} \frac{\delta \G}{\delta \phi_a} \nonumber \\                        &   & ~~~~~~ 
                        + \frac{\delta \G}{\delta \omega_a^*} \frac{\delta \G}{\delta \omega_a}
                        + \frac{\delta \G}{\delta \bar c^*} \frac{\delta \G}{\delta \bar c} 
                        + B_a \frac{\delta \G}{\delta \bar \omega_a} 
                        + v c \frac{\delta \G}{\delta X_1}
                   \Big )  = 0 \, .  \nonumber \\
\label{sec.3:ST}
\end{eqnarray}
We develop $\G$ according to the loop order as follows
\begin{eqnarray}
\G=\sum_{j=0}^\infty \hbar^{(j)} \G^{(j)} \, .
\label{sec.3:e6.new}
\end{eqnarray}
From eq.(\ref{sec.3:e5}) we get
\begin{eqnarray}
&& \frac{\delta \G^{(j)}}{\delta X_1} = \frac{1}{v} (\square + 2 \lambda v^2) \frac{\delta \G^{(j)}}{\delta \bar c^*} \, , ~~~~~ j \geq 1 
\label{sec.3:n1}
\end{eqnarray}
and therefore $\G^{(j)}$ depends on $X_1$ only via the combination
\begin{eqnarray}
\bar c^* + \frac{1}{v} (\square + 2 \lambda v^2) X_1 \, .
\label{sec.3:comb1}
\end{eqnarray}
From eq.(\ref{sec.3:e6}) we get
\begin{eqnarray}
&& \frac{\delta \G^{(j)}}{\delta X_2} = \frac{1}{v} (\square + 2 \lambda v^2) \frac{\delta \G^{(j)}}{\delta \bar c^*} \, , ~~~~~ j \geq 1 
\label{sec.3:n2}
\end{eqnarray}
which implies that $\G^{(j)}$ depends on $X_2$ only via
the combination 
\begin{eqnarray}
\bar c^* + \frac{1}{v} (\square + 2 \lambda v^2) X_2 \, .
\label{sec.3:comb2}
\end{eqnarray}
Eq.(\ref{sec.3:comb1}) together with eq.(\ref{sec.3:comb2})
yields that the dependence of $\G^{(j)}$ on $X_1,X_2$ is only via
\begin{eqnarray}
\widehat{\bar c^*} = \bar c^* + \frac{1}{v} (\square + 2 \lambda v^2) (X_1 + X_2) \, .
\label{sec.3:comb3}
\end{eqnarray} 
Moreover from eq.(\ref{sec.3:e1}) we have
\begin{eqnarray}
&& \frac{\delta \G^{(j)}}{\delta \bar \omega_a} = \partial^\mu \frac{\delta \G^{(j)}}{\delta A^{\mu*}_a}
   + \xi gv \frac{\delta \G^{(j)}}{\delta \phi_a^*} \, , ~~~~~~~~ j \geq 1 \, ,
\label{sec.3:n5}
\end{eqnarray}
i.e. $\G^{(j)}$ depends on $\bar \omega_a$ only via
the combinations
\begin{eqnarray}
\widehat{A^{*}_{a\mu}} = A^{*}_{a\mu} - \partial_\mu \bar \omega_a \, , ~~~~~
\widehat{\phi_a^*} = \phi_a^* + \xi gv \bar \omega_a \, .
\label{sec.3:n6}
\end{eqnarray}
From eqs.(\ref{sec.3:e2}) and (\ref{sec.3:e3})  we get
\begin{eqnarray}
\frac{\delta \G^{(j)}}{\delta \bar c} = 0 \, , ~~~~
\frac{\delta \G^{(j)}}{\delta c} = 0 \, , ~~~~~ j \geq 1 \, ,
\label{sec.3:n7}
\end{eqnarray}
and thus $\G^{(j)}$ does not depend on $\bar c, c$.
From eq.(\ref{sec.3:e4}) we obtain 
\begin{eqnarray}
\frac{\delta \G^{(j)}}{\delta B_a} = 0 \, , 
\label{sec.3:n8}
\end{eqnarray}
hence $\G^{(j)}$ does not depend on $B_a$.

Therefore we can restrict the analysis of the divergences of the theory to the 1-PI Green functions depending on $A_{a\mu}, \phi_a, \sigma, \omega_a,
A^*_{a\mu}, \phi_a^*, \sigma^*, \omega_a^*, \bar c^*$
(those depending on at least one of $X_1, X_2, \bar \omega_a$ can be obtained
by functional differentiation of eqs.(\ref{sec.3:n1}), (\ref{sec.3:n2}) and (\ref{sec.3:n5}) respectively).
In all these amplitudes $X_1$ and $X_2$ are exchanged within 1-PI graphs
in the combination $X=X_1 + X_2$. The latter is associated to
the propagator
\begin{eqnarray}
\Delta_{XX} & = & \Delta_{X_1 X_1} + \Delta_{X_2 X_2} =
                  -\frac{i}{p^2 - 2 \lambda v^2} 
                  +\frac{i}{p^2 - (2 \lambda + \frac{1}{\rho})v^2} \nonumber \\
            & = & \frac{iv^2}{\rho (p^2 - 2 \lambda v^2)(p^2 - (2 \lambda + \frac{1}{\rho})v^2)}  
\label{sec.3:n9}
\end{eqnarray}
which falls off for $p^2 \rightarrow \infty$ as $1/(p^2)^2$. This means that $X$ has UV dimension zero and thus the vertex in eq.(\ref{sec.3:13}) still obeys
the power-counting renormalizability bounds.
Moreover, since $A_{a\mu}, \phi_a, \sigma, \omega_a,
A^*_{a\mu}, \phi_a^*, \sigma^*, \omega_a^*, \bar c^*$ have positive dimension,
only a finite number of counterterms is needed in order to remove
all the divergences of the theory. 

\subsection{Structure of the counterterms}\label{sec:cts}

We assume that divergences have been recursively subtracted 
up to order $n-1$ in the loop expansion and that the ST identity
holds up to order $n$.
We assume as well 
that the set of functional equations (\ref{sec.3:e1})-(\ref{sec.3:e6})
is fulfilled up to order $n$.
The $n$-th order ST identity reads for the symmetrically regularized $n$-th order vertex functional
$\G^{(n)}_R$
\begin{eqnarray}
\delta \G_R^{(n)} = - \sum_{j=1}^{n-1} (\G^{(j)},\G^{(n-j)}) 
\label{sec3:t1}
\end{eqnarray}
where $\delta$ is the linearized ST operator
\begin{eqnarray}
\delta & = & \int d^4x \, \Big [ (D_\mu \omega)_a \frac{\delta}{\delta A_{a\mu}} - g \omega_a \phi_a \frac{\delta}{\delta \sigma} + g (\omega_a( \sigma + v) + \epsilon_{abc} \phi_b \omega_c) \frac{\delta}{\delta \phi_a} \nonumber \\
       &   & ~~~~~~ - \frac{1}{2} g f_{abc} \omega_b \omega_c \frac{\delta}{\delta \omega_a} 
             + B_a \frac{\delta}{\delta \bar \omega_a} + vc \frac{\delta}{\delta X_1} 
             + \Big ( \frac{1}{2} \sigma^2 + v \sigma + \frac{1}{2} \phi_a^2 - v X_2 \Big ) \frac{\delta}{\delta \bar c} \nonumber \\
       &   & ~~~~~~ + \frac{\delta \G^{(0)}}{\delta A_a^\mu} \frac{\delta}{\delta A^*_{a\mu}} 
             + \frac{\delta \G^{(0)}}{\delta \phi_a} \frac{\delta}{\delta \phi_a^*} 
             + \frac{\delta \G^{(0)}}{\delta \sigma} \frac{\delta}{\delta \sigma^*} 
             + \frac{\delta \G^{(0)}}{\delta \omega_a} \frac{\delta}{\delta \omega_a^*} 
             + \frac{\delta \G^{(0)}}{\delta \bar c} \frac{\delta}{\delta \bar c^*} \Big ]         
\label{sec3:t2}
\end{eqnarray}
and the bracket in the R.H.S. of eq.(\ref{sec3:t1}) is given by
\begin{eqnarray}
&& \!\!\!\!\!\!\!\! 
   (X,Y)  =  \int d^4x \, \Big [ \frac{\delta X}{\delta A^{*\mu}_a} \frac{\delta Y}{\delta A_{a\mu}} 
             + \frac{\delta X}{\delta \phi_a^*} \frac{\delta Y}{\delta \phi_a} 
             + \frac{\delta X}{\delta \sigma^*} \frac{\delta Y}{\delta \sigma} 
             + \frac{\delta X}{\delta \omega_a^*} \frac{\delta Y}{\delta \omega_a} 
             + \frac{\delta X}{\delta \bar c^*} \frac{\delta Y}{\delta \bar c} \Big ] \, . \nonumber \\
\label{sec3:t3}
\end{eqnarray}
Since the divergences have been recursively subtracted up to order $n-1$, the R.H.S. of eq.(\ref{sec3:t1})
is finite. Thus, as a consequence of eq.(\ref{sec3:t1}), the divergent part $\G^{(n)}_{R,div}$ of
$\G^{(n)}_R$ must obey the linearized ST identity
\begin{eqnarray}
\delta \G^{(n)}_{R,div} = 0 \, .
\label{sec3:t4}
\end{eqnarray}
In order to solve eq.(\ref{sec3:t4}) it is useful to decompose $\delta$
according to the degree induced by the counting operator
for the $\delta$-invariant variable $\widehat{\bar c^*}$ in eq.(\ref{sec.3:comb3}).
Then $\delta$ can be written as
\begin{eqnarray}
\delta = \delta_0 + \delta_1 \, , 
\label{sec3:t4.biss}
\end{eqnarray}
where $\delta_0$ preserves the number of $\widehat{\bar c^*}$'s and
$\delta_1$ increases it by one. The explicit action 
of the differentials $\delta_0,\delta_1$ on the variables
of the model is given in eqs.~(\ref{appB:3.5}) and (\ref{appB:4}).

The most general solution to eq.(\ref{sec3:t4}) of dimension $\leq 4$ and subject to the constraints
in (\ref{sec.3:n1}),(\ref{sec.3:n2}),(\ref{sec.3:n5}),
(\ref{sec.3:n7}) and (\ref{sec.3:n8})
is derived in Appendix ~\ref{appB}. It is given by
\begin{eqnarray}
\G^{(n)}_{R,div} & = &  d_1 \int d^4x \, G_{\mu\nu a} G^{\mu\nu}_a + d_2 \int d^4x \, (D_\mu \Phi)^\dagger (D^\mu \Phi)
             \nonumber \\
       &   &  +d_3 \int d^4x \, (\Phi^\dagger \Phi - \frac{v^2}{2}) + d_4 \int d^4x \, (\Phi^\dagger \Phi - \frac{v^2}{2})^2 \nonumber \\
       &   & +d_5 \int d^4x \, \delta_0 ( \widehat \phi^*_a \phi_a ) + 
              d_6 \int d^4x \, \delta_0 ( \sigma^* \sigma ) \nonumber \\
       &   & + d_7 \int d^4x \, \delta_0 (\widehat {A^*_{a\mu}} A_a^\mu) +
              d_8 \int d^4x \, \delta_0 (\omega_a^* \omega_a) 
             + d_9 \int d^4x \, \widehat{\bar c^*}  \nonumber \\
       &   & + d_{10} \int d^4x \,  ~ \widehat {\bar c^*} ( \Phi^\dagger \Phi - \frac{v^2}{2} )      
             + d_{11} \int d^4x \, \widehat {\bar c^*}^2 \nonumber \\
       &   & + \int d^4x \, \widehat {\bar c^*} [ (d_6 - d_5)  \sigma^2 
                                               + (d_6 - 2 d_5) v \sigma ]  \, ,
\label{sec3:div.1}
\end{eqnarray}
where $d_1, \dots, d_{11}$ parameterize the $n$-th loop overall local divergences.
After the recursive subtraction has been performed, the $n$-th order local divergences
$\G^{(n)}_{R,div}$ are removed by adding the $n$-th order counterterms $- \G^{(n)}_{R,div}$.
The ST identity is preserved by this subtraction. 

We notice that one can always add to the resulting
$n$-th order vertex functional 
$$\G^{(n)} = \G^{(n)}_R + ( -\G^{(n)}_{R,div})$$ a functional of the same form
as in eq.(\ref{appB:12}) with finite coefficients $a_1, \dots, a_{11}$ while preserving
the $n$-th order ST identity and the functional equations
(\ref{sec.3:e1})-(\ref{sec.3:e6}).
These ambiguities have to be fixed by a suitable choice of normalization conditions.
A convenient set of normalization conditions is given at
the end of Sect.~\ref{sec.6}.

\section{Physical Unitarity}\label{sec.6}

In this Section we address the issue of Physical Unitarity.
We first discuss the tree-level approximation and then move to the analysis of the renormalized theory.

\subsection{Tree-level}

The ST identity in eq.(\ref{sec.3:11}) yields by projection at order zero 
in powers of $c$ the following functional identity:
\begin{eqnarray}
\!\!\!\!\!\! {\cal S}_0(\G^{(0)}) & = & \int d^4x \, 
                   \Big ( \frac{\delta \G^{(0)}}{\delta A^{\mu*}_a} \frac{\delta \G^{(0)}}{\delta A_{a\mu}} 
                        + \frac{\delta \G^{(0)}}{\delta \sigma^*} \frac{\delta \G^{(0)}}{\delta \sigma}
                        + \frac{\delta \G^{(0)}}{\delta \phi^*_a} \frac{\delta \G^{(0)}}{\delta \phi_a} \nonumber \\                        &   & ~~~~~~ 
                        + \frac{\delta \G^{(0)}}{\delta \omega_a^*} \frac{\delta \G^{(0)}}{\delta \omega_a}
                        + B_a \frac{\delta \G^{(0)}}{\delta \bar \omega_a} 
                        \Big )  = 0 \, . 
\label{sec6:unit1}
\end{eqnarray}
Moreover, projection of eq.(\ref{sec.3:11}) at order one in powers of $c$ yields
\begin{eqnarray}
\!\!\!\!\!\! {\cal S}_1(\G^{(0)}) & = & \int d^4x \, 
                   \Big ( \frac{\delta \G^{(0)}}{\delta \bar c^*} \frac{\delta \G^{(0)}}{\delta \bar c} 
                        + v c \frac{\delta \G^{(0)}}{\delta X_1}
                   \Big )  = 0 \, .  
\label{sec6:unit2}
\end{eqnarray}
The functional identity in eq.(\ref{sec6:unit1}) is generated by the invariance of $\G^{(0)}$ under
the BRST differential $s_0$ in eq.(\ref{sec.3:BRST1}), the functional identity in eq.(\ref{sec6:unit2})
is generated by the invariance of $\G^{(0)}$ under the BRST differential $s_1$ in eq.(\ref{sec.3:BRST2}).

Correspondingly there are two conserved asymptotic charges $Q_0$ and $Q_1$ associated with 
eq.(\ref{sec6:unit1}) and (\ref{sec6:unit2}) respectively. They 
act as follows on the fields of the theory ($[ ~, ~ ]_+$ denotes the anticommutator):
\begin{eqnarray}
&& 
\!\!\!\!\!\!\!\!\!\!\!\!\!\!
[Q_0,A_{a\mu}] = \partial_\mu \omega_a \, , ~~~~ [Q_0, \omega_a]_+ = 0 \, , \nonumber \\
&& 
\!\!\!\!\!\!\!\!\!\!\!\!\!\!
[Q_0,\phi_a] = g v \omega_a \, , ~~~~ [Q_0,\bar \omega_a]_+ = B_a \, , ~~~~ [Q_0,B_a]=0 \, ,  \nonumber \\
&& 
\!\!\!\!\!\!\!\!\!\!\!\!\!\!
[Q_0,\sigma] = 0 \, , ~~~~~ [Q_0,X_1] = 0  \, , ~~~~ [Q_0,X_2]=0 \, , \nonumber  \\
&&
\!\!\!\!\!\!\!\!\!\!\!\!\!\!
[Q_0,\bar c]_+ = 0 \, , ~~~~ [Q_0,c]_+ = 0 \, ,
\label{sec.6:unit.1}
\end{eqnarray}
and
\begin{eqnarray}
&& 
\!\!\!\!\!\!\!\!\!\!\!\!\!\!
[Q_1,A_{a\mu}] = [Q_1,\phi_a]  = [Q_1, \sigma] = [Q_1, B_a] =  [Q_1, \omega_a]_+ =  [Q_1,\bar \omega_a]_+ = 0 \, ,  \nonumber \\
&& 
\!\!\!\!\!\!\!\!\!\!\!\!\!\!
[Q_1,X_1] =vc  \, , ~~~~ [Q_1,X_2]=0 \, ,  
\nonumber  \\
&&
\!\!\!\!\!\!\!\!\!\!\!\!\!\!
[Q_1,\bar c]_+ = v \sigma - v X_2 \, , ~~~~ [Q_1,c]_+ = 0 \, .
\label{sec.6:unit.1bis}
\end{eqnarray}

We characterize the physical Hilbert space ${\cal H}_{phys}$ as the space
\begin{eqnarray}
{\cal H}_{phys} = {\cal H}_0 \cap {\cal H}_1 \, ,
\label{sec.6:unit.2}
\end{eqnarray}
where
\begin{eqnarray}
{\cal H}_0 = \frac{\mbox{Ker } Q_0}{\mbox{Im } Q_0} ~~~~~ \mbox{and}  ~~~~~ {\cal H}_1 = \frac{\mbox{Ker } Q_1}{\mbox{Im } Q_1} \, .
\label{sec.6:unit.2.2}
\end{eqnarray}
That is, ${\cal H}_{phys}$ is the intersection of the quotient spaces \cite{Curci:1976yb}-\cite{Becchi:1985bd}
associated with the two conserved BRST charges $Q_0$ and $Q_1$.

In the sector spanned by $\sigma, X_1, X_2$
the mass eigenstates are $\sigma' = \sigma - X_1 - X_2$, $X_1$ and $X_2$. $\sigma'$ and $X_1$ have mass $p^2 = 2 \lambda v^2$,
$X_2$ has mass $p^2 = (2 \lambda + \frac{1}{\rho})v^2$.
$\bar c$ and $c$ have mass $p^2 = 2 \lambda v^2$.
$\bar \omega_a, \omega_a$ have mass $p^2 = \xi (gv)^2$, $\phi_a$ and the longitudinal 
component $\partial A_a$ of $A_{a\mu}$ have mass $p^2 = \xi (gv)^2$.

\medskip
We first construct ${\cal H}_0$. From eq.(\ref{sec.6:unit.1}) we 
see that ${\cal H}_0$ contains
$\bar c, c$, $X_1$, $X_2$ and $\sigma'$. Moreover the only  modes 
belonging to ${\cal H}_0$ 
in the sector spanned by $A_{a\mu}, \phi_a$, $B_a, \bar \omega_a$ and $\omega_a$ 
are the three transverse components
(in the four dimensional sense) of $A_{a \mu}$, i.e. those whose polarization vector
$\epsilon_\mu(p)$ fulfills
\begin{eqnarray}
\epsilon_\mu(p) p^\mu = 0 ~~~~~~ \mbox{at } p^2 = (M^{(0)}_A)^2 = (gv)^2 \, .
\label{sec.6:unit.2.bis}
\end{eqnarray}
In the above equation $M^{(0)}_{A}$ stands for the tree-level mass of $A_{a\mu}$.
Indeed we find (in the momentum space representation)
\begin{eqnarray}
[Q_0, \epsilon_\mu(p) A^\mu_a(p) ] = - i \epsilon_\mu(p) p^\mu \omega_a = 0 \, .
\label{sec.6:unit.new1}
\end{eqnarray}
Moreover $\bar \omega_a$ and $B_a$ are $Q_0$-doublets
\cite{Curci:1976yb}-\cite{Becchi:1985bd}, \cite{Piguet:1995er}-\cite{Quadri:2002nh}:
\begin{eqnarray}
[Q_0, \bar \omega_a]_+ = B_a \, , ~~~~ [Q_0, B_a] = 0 \, 
\label{sec.6:unit.3.ter}
\end{eqnarray}
and hence they are not in ${\cal H}_0$. 
The ghost $\omega_a$ is also paired into a $Q_0$-doublet with the longitudinal polarization $\rho_\mu(p)$
of $A_{a\mu}$ (i.e. such that $\rho_\mu(p) p^\mu = 1$ at 
$p^2 = \xi (M^{(0)}_A)^2$):
\begin{eqnarray}
[Q_0, \rho_\mu(p) A^\mu_a(p) ] = - i \rho_\mu(p)p^\mu \omega_a \, .
\label{sec.6:unit.3.quater}
\end{eqnarray}
From the above equation we see that $\omega_a$  and 
$\rho_\mu(p) A^\mu_a(p)$ are not in ${\cal H}_0$ . Finally 
$\phi_a$  does not belong to the kernel of $Q_0$ and 
thus it is outside ${\cal H}_0$.

\medskip
We now characterize  ${\cal H}_1$. Since by eq.(\ref{sec.6:unit.1bis}) 
\begin{eqnarray}
[Q_1, \sigma '] =- vc \, 
\label{sec.6:unit.3}
\end{eqnarray}
we get that $\sigma'$ is not in ${\cal H}_1$. By eq.(\ref{sec.6:unit.1bis}) we also see that 
 $X_1$ is not in ${\cal H}_1$ while $X_2$ belongs to ${\cal H}_1$.
For any finite value of $\rho$ the $Q_1$-invariant 
combination $\sigma' + X_1$ is $Q_1$-exact since
\begin{eqnarray}
[Q_1, \bar c]_+ = v \sigma - v X_2 = v (\sigma' + X_1) \, . 
\label{unit.2}
\end{eqnarray}
Therefore $\sigma' + X_1$ does not belong to ${\cal H}_1$.
From the above equation we also see that $\bar c$ is not in ${\cal H}_1$.
Furthermore $c$ is not in ${\cal H}_1$ since it forms a $Q_1$-doublet with $\frac{1}{v} X_1$. 
This implies that the only mode in ${\cal H}_1$ in the sector spanned by $X_1, X_2, \sigma, \bar c,c$
is $X_2$.
Its mass is given by
\begin{eqnarray}
m_{X_2} = (2 \lambda + \frac{1}{\rho}) v^2 \, .
\label{unit.3}
\end{eqnarray}
From eq.(\ref{sec.6:unit.1bis}) we get that $A_{a\mu}, \phi_a, \bar \omega_a, \omega_a, B_a$ are also in ${\cal H}_1$.

\medskip
By taking into account the above construction of ${\cal H}_0$ and ${\cal H}_1$ we conclude
according to eq.(\ref{sec.6:unit.2}) that ${\cal H}_{phys}$ is spanned by the transverse polarizations of
$A_{a\mu}$ in eq.(\ref{sec.6:unit.2.bis}) and by the scalar $X_2$.

\subsection{Higher orders}

The analysis of the physical states in the renormalized theory follows a similar path.
By eq.(\ref{sec.3:e3}) the ST identity in eq.(\ref{sec.3:ST}) can be projected at order
zero in powers of $c$ yielding
\begin{eqnarray}
\!\!\!\!\!\! {\cal S}_0(\G) & = & \int d^4x \, 
                   \Big ( \frac{\delta \G}{\delta A^{\mu*}_a} \frac{\delta \G}{\delta A_{a\mu}} 
                        + \frac{\delta \G}{\delta \sigma^*} \frac{\delta \G}{\delta \sigma}
                        + \frac{\delta \G}{\delta \phi^*_a} \frac{\delta \G}{\delta \phi_a} \nonumber \\                        &   & ~~~~~~ 
                        + \frac{\delta \G}{\delta \omega_a^*} \frac{\delta \G}{\delta \omega_a}
                        + B_a \frac{\delta \G}{\delta \bar \omega_a} 
                        \Big )  = 0 \, . 
\label{sec6:unit1.ren}
\end{eqnarray}
Moreover, the projection of eq.(\ref{sec.3:ST}) at order one in powers of $c$ gives
\begin{eqnarray}
\!\!\!\!\!\! {\cal S}_1(\G) & = & \int d^4x \, 
                   \Big ( \frac{\delta \G}{\delta \bar c^*} \frac{\delta \G}{\delta \bar c} 
                        + v c \frac{\delta \G}{\delta X_1}
                   \Big )  = 0 \, .  
\label{sec6:unit2.ren}
\end{eqnarray}
These are the renormalized ST identities associated with the BRST differentials $s_0$ and $s_1$ respectively.

By taking into account global SU(2) invariance and eq.(\ref{sec.3:e3})
we derive the action of the conserved asymptotic charges $Q_0$ and $Q_1$ associated with 
eq.(\ref{sec6:unit1.ren}) and (\ref{sec6:unit2.ren}) 
on the fields of the theory:
\begin{eqnarray}
&& 
\!\!\!\!\!\!\!\!\!\!\!\!\!\!
[Q_0,A_{a\mu}] = \G_{\omega_b A_{a\mu}^*} \omega_b \, , ~~~~ [Q_0, \omega_a]_+ = 0 \, , \nonumber \\
&& 
\!\!\!\!\!\!\!\!\!\!\!\!\!\!
[Q_0,\phi_a] = \G_{\omega_b \phi_a^*} \omega_b \, , ~~~~ [Q_0,\bar \omega_a]_+ = B_a \, , ~~~~ [Q_0,B_a]=0 \, ,  \nonumber \\
&& 
\!\!\!\!\!\!\!\!\!\!\!\!\!\!
[Q_0,\sigma] = 0 \, , ~~~~~ [Q_0,X_1] = 0  \, , ~~~~ [Q_0,X_2]=0 \, , \nonumber  \\
&&
\!\!\!\!\!\!\!\!\!\!\!\!\!\!
[Q_0,\bar c]_+ = 0 \, , ~~~~ [Q_0,c]_+ = 0 \, ,
\label{sec.6:unit.1.ren}
\end{eqnarray}
and
\begin{eqnarray}
&& 
\!\!\!\!\!\!\!\!\!\!\!\!\!\!
[Q_1,A_{a\mu}] = [Q_1,\phi_a] = [Q_1, \sigma] = [Q_1, B_a] = [Q_1, \omega_a]_+ =  [Q_1,\bar \omega_a]_+ = 0 \, ,  \nonumber \\
&& 
\!\!\!\!\!\!\!\!\!\!\!\!\!\!
[Q_1,X_1] =vc  \, , ~~~~ [Q_1,X_2]=0 \, , \nonumber  \\
&&
\!\!\!\!\!\!\!\!\!\!\!\!\!\!
[Q_1,\bar c]_+ = \G_{\sigma \bar c^*} \sigma +
 \G_{X_1 \bar c^*} X_1 +  \G_{X_2 \bar c^*} X_2 \, , ~~~~ [Q_1,c]_+ = 0 \, .
\label{sec.6:unit.1bis.ren}
\end{eqnarray}
The shorthand notations $\G_{\omega_b A_{a\mu}^*}$, $\G_{\omega_b \phi_a^*}$, $\G_{\sigma \bar c^*}$, $\G_{X_1 \bar c^*}$ and $\G_{X_2 \bar c^*}$  stand for the two-point 1-PI Green functions
\begin{eqnarray}
&& \!\!\!\!\!\!
\G_{\omega_b A_{a\mu}^*} = \left. \frac{\delta^2 \G}{\delta \omega_b(-p) \delta A_{a\mu}^*(p)} \right |_{\zeta = 0} \, , 
~~~~ \G_{\omega_b \phi_a^*} = \left. \frac{\delta^2 \G}{\delta \omega_b(-p) \delta \phi_a^*(p)} \right |_{\zeta = 0} \, , \nonumber \\
&& \!\!\!\!\!\! 
\G_{\sigma \bar c^*} = \left. \frac{\delta^2 \G}{\delta \sigma(-p) \delta \bar c^*(p)} \right |_{\zeta = 0} \,, ~~~~~~~~~~
\G_{X_1 \bar c^*} = \left. \frac{\delta^2 \G}{\delta X_1(-p) \delta \bar c^*(p)} \right |_{\zeta = 0} \,, \nonumber \\
&& ~~~~~~~~~~~~~~~~~~~~ \G_{X_2 \bar c^*} = \left. \frac{\delta^2 \G}{\delta X_2(-p) \delta \bar c^*(p)} \right |_{\zeta = 0} \, ,
\label{sec.6:shorthand}
\end{eqnarray}
where $\zeta$ is a collective notation for all the fields and external sources of the theory. It is also useful to introduce the scalar form factor $G(p^2$) for 
$\G_{\omega_b A_{a\mu}^*}$ by setting
\begin{eqnarray}
\G_{\omega_b A_{a\mu}^*} = i p_\mu \delta^{ab} G(p^2) \, .
\label{sec.6:formfactor}
\end{eqnarray}

\medskip
Again the physical Hilbert space ${\cal H}_{phys}$ is defined as the intersection of the quotient spaces ${\cal H}_0$ and ${\cal H}_1$
associated with the conserved charges $Q_0$ and $Q_1$.

\medskip
We study first  ${\cal H}_0 = \mbox{Ker } Q_0/ \mbox{Im } Q_0$. 
From eq.(\ref{sec.6:unit.1.ren}) we get that $\sigma, X_1, X_2$
$\bar c$ and $c$ belong to ${\cal H}_0$.
In the sector spanned by 
$A_{a\mu},\phi_a,B_a,\bar \omega_a,\omega_a$ 
the analysis proceeds as in the standard treatment 
given in \cite{Curci:1976yb}-\cite{Becchi:1985bd}.
From the first of eqs.(\ref{sec.6:unit.1.ren}) we obtain that 
the transverse polarizations $\epsilon_\mu(p)$ of $A_{a\mu}$
(i.e. those obeying
\begin{eqnarray}
\epsilon_\mu(p) p^\mu = 0 ~~~~~~ \mbox{at } p^2 = M_A^2  \, 
\label{sec.6:unit.2.bis.ren}
\end{eqnarray}
where $M_A^2$ is the renormalized mass of the gauge bosons $A_{a\mu}$)
are in ${\cal H}_0$. This follows since
\begin{eqnarray}
[Q_0, \epsilon_\mu(p) A^\mu_a(p) ] & = & 
- i \epsilon_\mu(p) \G_{\omega_b A^{*\mu}_a} \omega_b  \nonumber \\
& = & \epsilon_\mu(p) p^\mu ~ G(p^2) \omega_a
= 0 \, .
\label{sec.6:unit.new1.ren}
\end{eqnarray}
In the above equation we have used eq.(\ref{sec.6:formfactor}) and  eq.(\ref{sec.6:unit.2.bis.ren}).
Eq.(\ref{sec6:unit1.ren}) together with eqs.(\ref{sec.3:e1}) and (\ref{sec.3:e4}) ensures \cite{Curci:1976yb}-\cite{Becchi:1985bd} 
that the unphysical modes
described by $\partial A_a$, $\phi_a$, $\bar \omega_a$ and $\omega_a$ 
have a common mass $M_\xi$ located at the solution of the equation
\begin{eqnarray}
\G_{\omega_b \bar \omega_a} = i p^\mu \G_{\omega_b A^*_{\mu a}} + \xi gv \G_{\omega_b \phi^*_a} = 0 \, .
\label{sec.6:unit.new2.ren}
\end{eqnarray} 

The longitudinal polarization $\rho_\mu(p)$, obeying
\begin{eqnarray}
\rho_\mu(p) p^\mu =1 ~~~~ \mbox{at } p^2 = M_\xi^2 \, , 
\label{sec.6:nn1}
\end{eqnarray}
forms a $Q_0$-doublet with $\omega_a$:
\begin{eqnarray}
[Q_0, \rho_\mu(p) A^\mu_a(p) ] & = & 
- i \rho_\mu(p) \G_{\omega_b A^*_{\mu a}} \omega_b  \nonumber \\
& = & \rho_\mu(p) p^\mu ~ G(p^2) \omega_a \, .
\label{sec.6:nn2}
\end{eqnarray}
Thus $\rho_\mu(p) A^\mu_a(p)$ and $\omega_a$ do not belong to ${\cal H}_0$.
$\phi_a$ is not in the kernel of $Q_0$ and hence it is outside ${\cal H}_0$.
Finally $\bar \omega_a$ and $B_a$ form a $Q_0$-doublet
\begin{eqnarray}
[Q_0 , \bar \omega_a ]_+ = B_a 
\label{sec.6:nn3}
\end{eqnarray}
and consequently they are not in ${\cal H}_0$.
We conclude that ${\cal H}_0$ is spanned by $X_1,X_2,\bar c,c, \sigma$
and the three transverse polarizations of $A_{a\mu}$.

\medskip
The analysis of ${\cal H}_1 = \mbox{Ker } Q_1 / \mbox{Im } Q_1$ 
at the quantum level requires to discuss the mixing in the
sector spanned by $\sigma, X_1, X_2$.
The relevant two point functions are 
controlled by eqs.(\ref{sec.3:e5}) and (\ref{sec.3:e6}). One gets
\begin{eqnarray}
&& \G_{\sigma X_1} = \frac{1}{v} (-p^2 + 2 \lambda v^2) \G_{\sigma \bar c^*} \, , \label{unit:s1} \\
&& \G_{\sigma X_2} = \frac{1}{v} (-p^2 + 2 \lambda v^2) \G_{\sigma \bar c^*} \, , \label{unit:s2} \\
&& \G_{X_1 X_1} = \frac{1}{v} (-p^2 + 2 \lambda v^2) \G_{X_1 \bar  c^*} = 
                  \Big ( \frac{1}{v} \Big )^2 (-p^2 + 2 \lambda v^2)^2 \G_{\bar c^* \bar c^*} \, , \label{unit:s3} \\
&& \G_{X_1 X_2} = \frac{1}{v} (-p^2 + 2 \lambda v^2) \G_{X_2 \bar c^*} \nonumber \\
&& ~~~~~~~~~~ = \frac{1}{v} (-p^2 + 2 \lambda v^2)
                  \Big [ \frac{1}{v} (-p^2 + 2 \lambda v^2) \G_{\bar c^* \bar c^*} - v \Big ] \, , \label{unit:s4}
\\
&& \G_{X_2 X_2} = \frac{1}{v} (-p^2 + 2 \lambda v^2) \G_{X_2 \bar c^*} - \frac{v^2}{\rho}
                 + (p^2 - 2 \lambda v^2) \nonumber \\
&& ~~~~~~~~~ = \frac{1}{v} (-p^2 + 2 \lambda v^2) \Big [ \frac{1}{v} (-p^2 + 2 \lambda v^2) \G_{\bar c^* \bar c^*}
           -2 v \Big ] - \frac{v^2}{\rho} \,                                                     
\label{unit:s5}
\end{eqnarray}
where we have used the fact that, again as a consequence of eqs.(\ref{sec.3:e5}) and (\ref{sec.3:e6}),
\begin{eqnarray}
&& \G_{X_1 \bar c^*} = \frac{1}{v} (-p^2 + 2 \lambda v^2) \G_{\bar c^* \bar c^*} \, , \nonumber \\
&& \G_{X_2 \bar c^*} = \frac{1}{v} (-p^2 + 2 \lambda v^2) \G_{\bar c^* \bar c^*} - v \, .
\label{unit.6}
\end{eqnarray}
The determinant of the two-point function matrix $\G_2$ in the sector
spanned by $\sigma, X_1, X_2$ is
\begin{eqnarray}
\mbox{det } \G_2 = \frac{1}{\rho} (p^2 -2 \lambda v^2)^2
                   (\G^2_{\sigma \bar c^*} - (\rho + \G_{\bar c^* \bar c^*}) \G_{\sigma\sigma} ) = 0 \, .
\label{unit.8}
\end{eqnarray}
Therefore the masses of the particles in this sector are located at
\begin{eqnarray}
p^2 = 2 \lambda v^2 
\label{unit.8.1}
\end{eqnarray}
and at the solution of the equation
\begin{eqnarray}
\G^2_{\sigma \bar c^*}(p^2) - (\rho + \G_{\bar c^* \bar c^*}(p^2)) \G_{\sigma\sigma}(p^2) = 0 \, . 
\label{unit.8.2}
\end{eqnarray}
We denote the solution to eq.(\ref{unit.8.2}) by
\begin{eqnarray}
p^2 = \overline{M}^2 \, .
\label{unit.8.2.1}
\end{eqnarray}

We notice the appearance 
in eqs.(\ref{unit:s3})-(\ref{unit:s5}) of 
the combination  $(p^2 - 2 \lambda v^2)^2$.
Its coefficient must be zero in order to ensure
that the asymptotic states are described by pure Klein-Gordon fields
(no dipole components).
Remarkably, from eqs.(\ref{unit:s3})-(\ref{unit:s5}) this requirement 
can be fulfilled by imposing the single normalization condition
\begin{eqnarray}
\left . \G_{\bar c^* \bar c^*} \right |_{p^2 =2 \lambda v^2} = 0 \, .
\label{unit.9}
\end{eqnarray}
The above normalization condition is compatible with the symmetries
of the theory. It can be imposed order by order in the loop expansion 
by exploiting the $\delta$-invariant 
\begin{eqnarray}
\int d^4x \, \widehat {\bar c^*}^2 \, ,
\label{unit.10}
\end{eqnarray}
which can be freely added to the $n$-th order effective action while 
preserving all the functional identities of the model.

\medskip
Next we decompose the two point function $\G_{\sigma \sigma}$ into its tree-level contribution and the
quantum correction $\Sigma_{\sigma \sigma}$ as follows
\begin{eqnarray}
\G_{\sigma \sigma}(p^2) = p^2 - 2 \lambda v^2 + \Sigma_{\sigma \sigma}(p^2) \, .
\label{unit.11}
\end{eqnarray}
It is convenient to use the $\delta$-invariant 
\begin{eqnarray}
\int d^4x \, \Big ( \Phi^\dagger \Phi - \frac{v^2}{2} \Big )^2  
\label{unit.12}
\end{eqnarray}
in order to enforce recursively, order by order in the loop expansion,
the normalization condition
\begin{eqnarray}
\left . \Sigma_{\sigma \sigma} \right |_{p^2 =2 \lambda v^2} = 0 \, .
\label{unit.13}
\end{eqnarray}

\medskip
The analysis of the states spanned by $\sigma, X_1, X_2$ can be done
by studying the eigenstates of the two-point matrix
\begin{eqnarray}
\G_2 = \pmatrix{\G_{\sigma\sigma} & \G_{\sigma X_1} & \G_{\sigma X_2} \cr
                \G_{X_1 \sigma}   & \G_{X_1 X_1}    & \G_{X_1 X_2} \cr
                \G_{X_2 \sigma}   & \G_{X_2 X_1}    & \G_{X_2 X_2} }
\label{sec.6:2point}
\end{eqnarray}
at $p^2 = 2 \lambda v^2$ and at $p^2 = \overline{M}^2$ respectively.

We first describe the asymptotic states at $p^2 = 2\lambda v^2$.
We introduce a vector $\underline{\varphi}_\sharp$ collecting the fields
$\sigma, X_1, X_2$ (at $p^2 =2 \lambda v^2$) by setting
\begin{eqnarray}
^T \underline{\varphi}_\sharp = ( \sigma_\sharp , {X_1}_\sharp , {X_2}_\sharp ) \, .
\label{sec6:as1}
\end{eqnarray}
The subscript $\sharp$ means that $\sigma, X_1, X_2$ are taken at $p^2 =2 \lambda v^2$.
The solutions of the equation
\begin{eqnarray}
\left . \G_2 \right |_{p^2 = 2 \lambda v^2} \underline{u} = 0 \, , 
\label{sec6:as2}
\end{eqnarray}
where we have set $^T \underline{u} = (u_\sigma, u_{X_1}, u_{X_2})$,  parameterize the asymptotic
states at $p^2 =2 \lambda v^2$ on the basis spanned by the components of $\varphi_\sharp$.
The field corresponding to the vector $\underline{u}$ 
is thus
\begin{eqnarray}
\!\!\!\!\!\!\!\!\!\!\!\!\!\!
\varphi_\sharp(\underline{u}) & = & \underline{u} \cdot \underline{\varphi}_\sharp \nonumber \\
                       & = & u_\sigma \sigma_\sharp + u_{X_1} {X_1}_\sharp 
                           + u_{X_2} {X_2}_\sharp \, .
\label{sec6:a3}
\end{eqnarray}
From eqs.(\ref{unit:s1})-(\ref{unit:s5}) and by taking into account
eq.(\ref{unit.9}) and eq.(\ref{unit.13}) we get that there are
two independent solutions to eq.(\ref{sec6:as2}):
\begin{eqnarray}
^T \underline{u}_1 = (1,0,0) \, , ~~~~~ ^T \underline{u}_2 = (0,1,0) \, 
\label{sec6:a4}
\end{eqnarray}
so that 
\begin{eqnarray}
\varphi_\sharp(\underline{u}_1) = \sigma_\sharp \, , ~~~~ \varphi_\sharp(\underline{u}_2) = {X_1}_\sharp \, .
\label{sec6:a5.bis}
\end{eqnarray}
$\underline{u}_1$ and $\underline{u}_2$ allow to introduce a projector
$\Pi_{2\lambda v^2}$ on the mass eigenstates at $p^2 = 2 \lambda v^2$.
$\Pi_{2\lambda v^2}$ acts on any vector $^T \underline{w} = (w_1, w_2, w_3)$ 
as follows:
\begin{eqnarray}
\Pi_{2\lambda v^2}(\underline{w}) = (\underline{u}_1 \cdot \underline{w}) \underline{u}_1 +  (\underline{u}_2 \cdot \underline{w}) \underline{u}_2 \, .
\label{sec6:a5}
\end{eqnarray}
Correspondingly the action on  $\varphi_\sharp(\underline{w})$ is given by 
\begin{eqnarray}
\!\!\!\!\!\!\!\!
\Pi_{2\lambda v^2}(\varphi_\sharp(\underline{w})) = (\underline{u}_1 \cdot \underline{w}) \varphi_\sharp(\underline{u}_1) +  (\underline{u}_2 \cdot \underline{w}) \varphi_\sharp(\underline{u}_2) = w_1 \sigma_\sharp + w_2 {X_1}_\sharp \, .
\label{sec6:a6}
\end{eqnarray}

\medskip
From eq.(\ref{sec.6:unit.1bis.ren}) we see that $\varphi_\sharp(\underline{u}_2) = {X_1}_\sharp$
does not belong to ${\cal H}_1$ while $\varphi_\sharp(\underline{u}_1) = \sigma_\sharp$ does.
Moreover from eq.(\ref{sec.6:unit.1bis.ren}) we also obtain that 
$c$ is not in ${\cal H}_1$ since it forms a $Q_1$-doublet with $\frac{1}{v} {X_1}_\sharp$.
Furthermore $\sigma_\sharp$ is $Q_1$-exact also at the quantum level. 
Indeed from eq.(\ref{sec.6:unit.1bis.ren}) the action of $Q_1$ on
$\bar c$ reads
\begin{eqnarray}
[Q_1, \bar c]_+ & = & \G_{\sigma \bar c^*} \sigma_\sharp + \G_{X_1 \bar c^* } {X_1}_\sharp
             +\G_{X_2 \bar c^*} {X_2}_\sharp \, .
\label{unit.17}             
\end{eqnarray}
Since $\bar c$ has support at $p^2=2\lambda v^2$, we need to apply 
the operator $\Pi_{2\lambda v^2}$ to the R.H.S. in order to project it
on the subspace of asymptotic states at $p^2 =2 \lambda v^2$.
By eq.(\ref{sec6:a6}) we obtain
\begin{eqnarray}
[Q_1, \bar c]_+ & = & \Pi_{2\lambda v^2} (\G_{\sigma \bar c^*} \sigma_\sharp + \G_{X_1 \bar c^* } {X_1 }_\sharp
             +\G_{X_2 \bar c^*} {X_2}_\sharp) \nonumber \\
             & = & \G_{\sigma \bar c^*} \sigma_\sharp + \G_{X_1 \bar c^* } {X_1}_\sharp \, .
\label{unit.17.1}             
\end{eqnarray}
The second term in the second line of the above equation vanishes at $p^2 = 2\lambda v^2$ as a consequence of the first of eqs.(\ref{unit.6}).
Then one is left with
\begin{eqnarray}
[Q_1,\bar c]_+ & = & \G_{\bar c^* \sigma} \sigma_\sharp \, .
\label{unit.18}
\end{eqnarray}
Eq.(\ref{unit.18}) implies that $\sigma_\sharp$ is $Q_1$-exact provided
that 
\begin{eqnarray}
\left . \G_{\bar c^* \sigma} \right |_{p^2 = 2 \lambda v^2} \neq 0
\label{unit.19}
\end{eqnarray}
If eq.(\ref{unit.19}) is fulfilled, $\sigma$ does not belong to ${\cal H}_1$.
We notice that the condition in eq.(\ref{unit.19}) is verified at tree-level since 
$$\G^{(0)}_{\bar c^* \sigma} = v$$
and can be recursively preserved at the quantum level by making use of the $\delta$-invariant
\begin{eqnarray}
\int d^4x ~ \widehat{\bar c^*} \Big ( \Phi^\dagger \Phi - \frac{v^2}{2} \Big ) \, .
\label{unit.20}
\end{eqnarray}
Moreover, eq.(\ref{unit.19}) together with eq.(\ref{unit.13})
implies that the solution of eq.(\ref{unit.8.2})
cannot coincide with $p^2 =2 \lambda v^2$.
This implies that the solution of 
\begin{eqnarray}
\left . \G_2 \right |_{p^2 =\overline{M}^2} \underline{\tilde u} = 0 \, .
\label{sec6:asympt1}
\end{eqnarray}
at $p^2 = \overline{M}^2$ (asymptotic state at $p^2 = \overline{M}^2$) 
is $Q_1$-invariant. 
This can be proven as follows.
We denote by $\sigma_\flat$, ${X_1}_\flat$,
${X_2}_\flat$ the fields $\sigma, X_1, X_2$ at $p^2 = \overline{M}^2$.
Then the solution to eq.(\ref{sec6:asympt1}) is associated to the
field
\begin{eqnarray}
\varphi_\flat(\underline{\tilde u}) = {\tilde u}_\sigma \sigma_\flat + {\tilde u}_{X_1} {X_1}_\flat
+ {\tilde u}_{X_2} {X_2}_\flat \, .
\label{sec6:asympt2}
\end{eqnarray}
This is $Q_1$-invariant, since by eq.(\ref{sec.6:unit.1bis.ren}) 
$[Q_1, \sigma_\flat ] =0$, $[Q_1, {X_2}_\flat]=0$ and also
$[Q_1, {X_1}_\flat] =0$, due to the
fact that  by eq.(\ref{sec.3:6}) $c$ has support at $p^2 =2 \lambda v^2$.
It cannot be $Q_1$-exact since the only scalar $G$-singlet field
with negative ghost number is $\bar c$, which by eq.(\ref{sec.3:e3})
has support at $p^2 = 2 \lambda v^2$.
$\varphi_\flat(\underline{\tilde u})$ in eq.(\ref{sec6:asympt2}) is
the physical mode which is described at tree-level by the
field~$X_2$.

From eq.(\ref{sec.6:unit.1bis.ren}) we also see that $A_{a\mu}, \phi_a, B_a$,
$\omega_a, \bar \omega_a$ are in ${\cal H}_1$. Therefore
${\cal H}_1$ is spanned by $A_{a\mu}, \phi_a, B_a$,
$\omega_a, \bar \omega_a$ and the mode in eq.(\ref{sec6:asympt2}).

\medskip
By taking into account the above characterization of ${\cal H}_0$ and
${\cal H}_1$ we conclude that the space ${\cal H}_{phys}$ 
in eq.(\ref{sec.6:unit.2})
contains the three transverse polarization modes
of the gauge field $A_{a\mu}$ and a scalar particle 
with mass $p^2 = \overline{M}^2$ given by the solution of eq.(\ref{unit.8.2}).

\medskip
At this point we are in a position to provide the physical interpretation
of the parameters $a_1,\dots,a_{11}$ in eq.(\ref{appB:12}).
$a_1$ is associated with the finite 
renormalization of the gauge coupling constant,
$a_2$ with that of the mass of the non-abelian gauge bosons.
$a_3$ has to be used to impose the normalization condition
(absence of $\sigma$ tadpole)
\begin{eqnarray}
\left . \frac{\delta \G}{\delta \sigma} \right |_{\zeta = 0} (p^2=0)= 0 \, .
\label{sec6:physparam1}
\end{eqnarray}
Analogously $a_9$ allows to set the normalization condition
(absence of $X_1$ and $X_2$ tadpoles)
\begin{eqnarray}
\left . \frac{\delta \G}{\delta X_1} \right |_{\zeta = 0} (p^2=0) = 
\left . \frac{\delta \G}{\delta X_2} \right |_{\zeta = 0} (p^2=0) = 0 \, .
\label{sec6:physparam2}
\end{eqnarray}
$a_4$ is associated with the normalization condition on
 the two-point function $\G_{\sigma \sigma}$ at zero momentum and is used
to enforce eq.(\ref{unit.13}). $a_5,a_6,a_7,a_8$ are associated
to finite field redefinitions of $\phi_a, \sigma, A_{a\mu}, \omega_a$
respectively.  
By eq.(\ref{unit.8.2}) 
$a_{10}$ controls the finite renormalization of the mass of
the physical scalar mode.
Finally the freedom on the choice of 
$a_{11}$ is used in order to impose eq.(\ref{unit.9}), which
guarantees the absence of dipole fields in the asymptotic states
in the sector spanned by $\sigma, X_1,X_2$.

\section{Conclusions}\label{sec.7}

A polynomial formulation of the St\"uckelberg mechanism has been derived by making use
of an off-shell nilpotent BRST symmetry.
This symmetry is related to a natural abelian 
embedding of the St\"uckelberg action.
The antighost field of the U(1) symmetry
is responsible for the implementation of the St\"uckelberg constraint. 
Moreover we have shown that a mass term for the additional U(1)
gauge connection $B_\mu$ can be introduced
in a BRST-invariant way. 

We have proven that for the gauge group SU(2) 
a composite vector field transforming as a connection under
the BRST differential (but not under the SU(2) gauge
transformations) can be obtained by using the abelian antighost field. 
This allows us to generate a new polynomial BRST-invariant
but not gauge-invariant mass term for the non-abelian gauge fields.
We have given a sufficient condition for the
existence of this type of mass term for a general 
gauge group $G$.

The abelian embedded St\"uckelberg model discussed in this paper is not power-counting renormalizable.
We have shown that there is a natural theory which extends it to
a power-counting renormalizable model. The resulting theory is physically
unitary and contains in the physical sector the 
three physical polarizations of the massive gauge fields
as well as a physical  scalar particle.

The existence of the conserved charge $Q_0$ 
shows that the spontaneous symmetry breaking mechanism 
is implemented in the model.
Since the theory is power-counting renormalizable, one could conjecture that
it is physically equivalent to the Higgs model,
 i.e. that it yields the same physical S-matrix elements.
The check of the conjectured
equivalence in the full perturbative expansion is an interesting question
which deserves to be further investigated.

\section*{Acknowledgments}

Useful discussions with  R.~Ferrari and A.~A.~Slavnov are gratefully
acknowledged. 

\appendix
\section{Propagators}\label{appA}

By setting $B'_a = B_a - \frac{1}{\xi} (\partial A_a + \xi gv \phi_a)$ the propagators for $B'_a$, $A_{a \mu}$
and $\phi_a$ are diagonal:
\begin{eqnarray}
&& \Delta_{B'_a B'_b} = \frac{i}{\xi} \delta_{ab}  \, , ~~~~
\Delta_{\phi_a \phi_b} = \frac{i}{p^2 - \xi (gv)^2} \delta_{ab} \, , \nonumber \\
&& \Delta_{A_{a\mu} A_{b\nu}} = i \Big ( \frac{1}{-p^2 + (gv)^2} T^{\mu\nu} + 
                                         \frac{1}{-\frac{p^2}{\xi} + (gv)^2} \Big ) \delta_{ab} \, . 
\label{appA:1}
\end{eqnarray}
Moreover we set $\sigma = \sigma' + X_1 + X_2$.
Then the propagators for $X_1, X_2$ and $\sigma'$ are 
\begin{eqnarray}
&& \Delta_{\sigma' \sigma'} = \frac{i}{p^2 - 2 \lambda v^2} \, , \nonumber \\
&& \Delta_{X_1 X_1} = -\frac{i}{p^2 - 2 \lambda v^2} \, , \nonumber \\
&& \Delta_{X_2 X_2} = \frac{i}{p^2 - (2 \lambda + \frac{1}{\rho} ) v^2} \, .
\label{appA:2} 
\end{eqnarray}

In the ghost sector
\begin{eqnarray}
\Delta_{\bar \omega_a \omega_b} = \frac{i}{-p^2 + \xi (gv)^2} \delta_{ab} \, , ~~~~
\Delta_{\bar c c} = \frac{i}{p^2 - 2 \lambda v^2} \, .
\label{appA:3}
\end{eqnarray}

The remaining off-diagonal mixed propagators are all zero.

\section{Analysis of the cohomology of $\delta$ in the action-like sector}\label{appB}

The nilpotent linearized ST operator $\delta$ acts as follows on the fields and the antifields of the model:
\begin{eqnarray}
&& \delta A_{a\mu} = (D_\mu \omega)_a \, , ~~~~ \delta \sigma = - g \omega_a \phi_a \, , ~~~~
\delta \phi_a  = g (\omega_a (\sigma + v) + \epsilon_{abc} \phi_b \omega_c) \, , \nonumber \\
&& \delta X_1 = vc \, , ~~~~ \delta X_2 = 0 \, , ~~~~ \delta \bar c = \frac{1}{2} \sigma^2 + v \sigma + \frac{1}{2} \phi_a^2 -v X_2 \, , \nonumber \\
&& \delta c = 0 \, , ~~~~ \delta \bar \omega_a = B_a \, , ~~~~ \delta B_a = 0 \, , 
~~~~ \delta \omega_a = - \frac{1}{2} g f_{abc} \omega_b \omega_c \, , \nonumber \\
&& \delta \widehat{A_{a\mu}^*} = \frac{\delta \G_0}{\delta A_{a\mu}} \, , ~~~~
   \delta \sigma^* = \frac{\delta \G_0}{\delta \sigma} \, , ~~~~
   \delta \widehat \phi^*_a = \frac{\delta \G_0}{\delta \phi_a} \, , ~~~~
   \delta \widehat{\bar c^*} = 0 \, , \nonumber \\
&& \delta \omega_a^* =  \frac{\delta \G_0}{\delta \omega_a} 
\label{appB:1}
\end{eqnarray}
where $\G_0$ is given by
\begin{eqnarray}
\G_0 & = & S_\lambda 
\nonumber \\
                    &   &     + \int d^4x \, \Big ( \widehat{A_{a\mu}^*} (D^\mu \omega)_a 
  +\sigma^* (-g \omega_a \phi_a)
           +g \widehat \phi^*_a (\omega_a (\sigma + v) + \epsilon_{abc} \phi_b \omega_c ) \nonumber \\
     &   & ~~~~~~~~~~ - \frac{1}{2} \omega_a^* g f_{abc} \omega_b \omega_c + \widehat{\bar c^*} (\frac{1}{2} \sigma^2 
              + v \sigma + \frac{1}{2} \phi_a^2 - v X_2) \Big ) \, . 
\label{appB:2}                                                           
\end{eqnarray}
It is convenient to decompose $\delta$ according to the degree induced by the counting operator
for $\widehat{\bar c^*}$:
\begin{eqnarray}
\delta = \delta_0 + \delta_1 \, , 
\label{appB:3}
\end{eqnarray}
where $\delta_0$ preserves the number of $\widehat{\bar c^*}$'s and $\delta_1$ increases it by one.
$\delta_0$ is given by
\begin{eqnarray}
&& \delta_0 A_{a\mu} = (D_\mu \omega)_a \, , ~~~~ \delta_0 \sigma = - g \omega_a \phi_a \, , ~~~~
\delta_0 \phi_a  = g (\omega_a (\sigma + v) + \epsilon_{abc} \phi_b \omega_c) \, , \nonumber \\
&& \delta_0 X_1 = vc \, , ~~~~ \delta_0 X_2 = 0 \, , ~~~~ \delta_0 \bar c = \frac{1}{2} \sigma^2 + v \sigma + \frac{1}{2} \phi_a^2 -v X_2 \, , \nonumber \\
&& \delta_0 c = 0 \, , ~~~~ \delta_0 \bar \omega_a = B_a \, , ~~~~ \delta_0 B_a = 0 \, , 
~~~~ \delta_0 \omega_a = - \frac{1}{2} g f_{abc} \omega_b \omega_c \, , \nonumber \\
&& \delta_0 \widehat{A_{a\mu}^*} = \frac{\delta \G_0}{\delta A_{a\mu}} \, , ~~~~
   \delta_0 \sigma^* = \left . \frac{\delta \G_0}{\delta \sigma} \right |_{\widehat{\bar c^*} = 0} \, , ~~~~
   \delta_0 \widehat \phi^*_a = \left . \frac{\delta \G_0}{\delta \phi_a} \right |_{\widehat{\bar c^*} = 0} \, , ~~~~
   \delta_0 \widehat{\bar c^*} = 0 \, , \nonumber \\
&& \delta_0 \omega_a^* =  \frac{\delta \G_0}{\delta \omega_a} \, .
\label{appB:3.5}
\end{eqnarray}
$\delta_1$ is zero on all variables but $\sigma^*$ and $\widehat \phi^*_a$:
\begin{eqnarray}
\delta_1 \sigma^* = (\sigma + v) \widehat{\bar c^*} \, , ~~~~ 
\delta_1 \widehat \phi^*_a = \phi_a \widehat{\bar c^*} \, .
\label{appB:4}
\end{eqnarray}
We now derive the most general solution to the equation
\begin{eqnarray}
\delta \Sigma = 0 
\label{appB:5}
\end{eqnarray}
where $\Sigma$ is at most of dimension $4$ in the fields, the antifields and their derivatives and fulfills the same identities as $\G^{(j)}$ 
in eqs.(\ref{sec.3:n1}), (\ref{sec.3:n2}), (\ref{sec.3:n5}),
(\ref{sec.3:n7}) and (\ref{sec.3:n8}).
Since $\widehat{\bar c^*}$ has dimension two, the expansion of $\Sigma$ in powers of
$\widehat{\bar c^*}$ stops at the second order term
\begin{eqnarray}
\Sigma = \Sigma_0 + \Sigma_1 + \Sigma_2 
\label{appB:6}
\end{eqnarray}
where $\Sigma_j$ contains $j$ $\widehat{\bar c^*}$'s. Thus eq.(\ref{appB:5}) is equivalent to
the coupled set of equations
\begin{eqnarray}
&& \delta_0 \Sigma_0 = 0 \, , \nonumber \\
&& \delta_0 \Sigma_1 + \delta_1 \Sigma_0 = 0 \, , \nonumber \\
&& \delta_0 \Sigma_2 + \delta_1 \Sigma_1 = 0 \, .
\label{appB:7}
\end{eqnarray}
The solution to the first of the above equations is known  \cite{Piguet:1995er}, \cite{Barnich:2000zw}, \cite{Barnich:1994mt}-\cite{Barnich:1994ve} and can be written 
in terms of eight independent parameters $a_1 , \dots , a_8$
\begin{eqnarray}
\Sigma_0 & = & a_1 \int d^4x \, G_{\mu\nu a} G^{\mu\nu}_a + a_2 \int d^4x \, (D_\mu \Phi)^\dagger (D^\mu \Phi)
             \nonumber \\
       &   &  +a_3 \int d^4x \, (\Phi^\dagger \Phi - \frac{v^2}{2}) + a_4 \int d^4x \, (\Phi^\dagger \Phi - \frac{v^2}{2})^2 \nonumber \\
       &   & +a_5 \int d^4x \, \delta_0 ( \widehat \phi^*_a \phi_a ) + 
              a_6 \int d^4x \, \delta_0 ( \sigma^* \sigma ) \nonumber \\
       &   & + a_7 \int d^4x \, \delta_0 (\widehat {A^*_{a\mu}} A_a^\mu) +
              a_8 \int d^4x \, \delta_0 (\omega_a^* \omega_a) \, . 
\label{appB:8}
\end{eqnarray}
Evaluation of $\delta_1 \Sigma_0$ gives
\begin{eqnarray}
\delta_1 \Sigma_0 & = & \int d^4x \, \widehat {\bar c^*} [ 2 (a_6 - a_5) g \sigma \omega_a \phi_a
                                    + (a_6  - 2 a_5) gv \omega_a \phi_a ] \nonumber \\
                  & = & \delta_0 ( \int d^4x \, ( - \widehat {\bar c^*} [ (a_6 - a_5) \sigma^2 
                                               + (a_6 - 2 a_5) v \sigma ] ) ) \, .
\label{appB:9}
\end{eqnarray}
Therefore the second of eqs.(\ref{appB:7}) is solved by
\begin{eqnarray}
\Sigma_1 & = & \int d^4x \, \widehat {\bar c^*} [ (a_6 - a_5)  \sigma^2 
                                               + (a_6 - 2 a_5) v \sigma ]  \nonumber \\
         &   & + a_9 \int d^4x \, ~\widehat {\bar c^*}
          + a_{10} \int d^4x \,  ~ \widehat {\bar c^*} ( \Phi^\dagger \Phi - \frac{v^2}{2} ) \, ,                   \label{appB:10}                
\end{eqnarray}
where the terms in the second line of eq.(\ref{appB:10}) are $\delta_0$-invariant.
$a_9, a_{10}$ are free parameters. Obviously $\delta_1 \Sigma_1 = 0$ and thus
the last of eqs.(\ref{appB:7}) reduces to $\delta_0 \Sigma_2 =0$.
By power-counting 
\begin{eqnarray}
\Sigma_2 = a_{11} \int d^4x \, \widehat {\bar c^*}^2 
\label{appB:11}
\end{eqnarray}
where $a_{11}$ is again a free parameter. Finally we get that the most general solution to eq.(\ref{appB:5}), compatible
 with eqs.(\ref{sec.3:n1}), (\ref{sec.3:n2}), (\ref{sec.3:n5}),
(\ref{sec.3:n7}) and (\ref{sec.3:n8}), is
\begin{eqnarray}
\Sigma & = &  a_1 \int d^4x \, G_{\mu\nu a} G^{\mu\nu}_a + a_2 \int d^4x \, (D_\mu \Phi)^\dagger (D^\mu \Phi)
             \nonumber \\
       &   &  +a_3 \int d^4x \, (\Phi^\dagger \Phi - \frac{v^2}{2}) + a_4 \int d^4x \, (\Phi^\dagger \Phi - \frac{v^2}{2})^2 \nonumber \\
       &   & +a_5 \int d^4x \, \delta_0 ( \widehat \phi^*_a \phi_a ) + 
              a_6 \int d^4x \, \delta_0 ( \sigma^* \sigma ) \nonumber \\
       &   & + a_7 \int d^4x \, \delta_0 (\widehat {A^*_{a\mu}} A_a^\mu) +
              a_8 \int d^4x \, \delta_0 (\omega_a^* \omega_a)
             + a_9   \int d^4x \,  ~ \widehat {\bar c^*} \nonumber \\
       &   & + a_{10} \int d^4x \,  ~ \widehat {\bar c^*} ( \Phi^\dagger \Phi - \frac{v^2}{2} )      
             + a_{11} \int d^4x \, \widehat {\bar c^*}^2 \nonumber \\
       &   & + \int d^4x \, \widehat {\bar c^*} [ (a_6 - a_5)  \sigma^2 
                                               + (a_6 - 2 a_5) v \sigma ]  \, .
\label{appB:12}
\end{eqnarray}
We notice that the fact that 
the dependence of $\Sigma$ on $X_2$ is only via the combination $\widehat{\bar c^*}$ prevents 
the appearance of further $\delta$-invariants with dimension $\leq 4$ like
\begin{eqnarray}
\!\!\!\!\!
\int d^4x \, X_2 (\Phi^\dagger \Phi - \frac{v^2}{2} ) \, , ~~~~
\int d^4x \, X_2^2 (\Phi^\dagger \Phi - \frac{v^2}{2} ) \, , ~~~~
\int d^4x \, \delta (\sigma^* X_2) \, .
\label{appB:13}
\end{eqnarray}

\end{document}